\documentclass[12pt]{article}

\usepackage[T1]{fontenc}
\usepackage{textcomp}
\usepackage{mathptmx}

\begin{document}
\title{Problems in application of LDPC codes to information reconciliation in quantum key distribution protocols\thanks{%
IEICE Technical Report (ISSN 0913-5685), IT2009-41, Sept.\ 2009.
The author welcomes comments from readers.}}
\author{Ryutaroh Matsumoto\\
Dept.\ of Communications and Integrated Systems\\
Tokyo Institute of Technology, 152-8550 Japan\\
Email: ryutaroh@rmatsumoto.org}
\date{September 2009}
\maketitle
\begin{abstract}
The information reconciliation in a quantum key distribution protocol
can be studied separately from other steps in the protocol.
The problem of information reconciliation can be reduced to that
of distributed source coding. Its solution by LDPC codes is reviewed.
We list some obstacles preventing the LDPC-based distributed source coding
from becoming a more favorable alternative to
the Cascade protocol for information
reconciliation in quantum key distribution protocols.
This exposition does not require knowledge of the quantum theory.
\end{abstract}

\section{Introduction}
The quantum key distribution (QKD) protocol invented in
\cite{bennett84} is one of technologies nearest to practical
realization among various quantum information processing technologies.
The goal of a QKD protocol is to share a common random string,
called \emph{key},
between two legitimate users Alice and Bob secretly from
the eavesdropper Eve. Alice and Bob can use an authenticated
public classical channel between them to achieve the goal, but Eve can
see all the contents in the public channel.
In addition to this classical channel, there is a quantum channel
between Alice and Bob over which quantum objects are transmitted
from Alice to Bob. Observe that this is a quantum extension of
the model CW introduced by Ahlswede
and Csisz\'ar \cite[Section 9.2]{liang09}
with the classical noisy channel replaced by the quantum noisy one.

As categorized in \cite{renner05}, a QKD protocol can usually be
divided into four steps:
\begin{description}
\item[1. Quantum transmission and reception:]
Alice transmits randomly chosen quantum objects to Bob.
Bob measures received objects by a randomly chosen measurement method.
After this step, Alice and Bob have classical bits of the same length.
The remaining steps in a QKD protocol are purely classical information
processing, and all the processed data are classical.
\item[2. Channel parameter estimation:]
Alice and Bob publicly announce parts of transmitted objects and
measurement outcomes. From announced data, they estimate the channel
parameters between them. Usually, part of parameters remains unknown.
Remaining parts of Alice and Bob's bits are used for generating 
secret key.

The surprising feature of the quantum theory is that
(quantum counterpart of) the joint probability distribution
among Alice, Bob and Eve can be determined from the channel
parameter only between Alice and Bob, which cannot be done
within the classical secret key agreement.
\item[3. Information reconciliation:]
Alice and Bob make their bits identical by conversation over
the public channel.
\item[4. Privacy amplification:]
Alice and Bob shorten their bits by multiplying a binary
matrix to their identical bits. The resulting shortened bits
are almost statistically independent of all the information
possessed by Eve, which includes the conversation between
Alice and Bob over the public channel.
\end{description}
Note that the third and fourth steps are essentially
the same as the information theoretically secure
key agreement introduced by Maurer, Ahlswede,
and Csisz\'ar \cite[Chapter 9]{liang09}.
Thus, many parts of this exposition are also relevant to
the information theoretically secure
key agreement.

Traditional security proofs for QKD protocols,
for example \cite{shor00}, combines the information reconciliation
and the privacy amplification. Because of that,
we could not study the information reconciliation in QKD protocols
separately from the privacy amplification,
for example, we could not investigate what kind of the information
reconciliation was suitable without considering the privacy
amplification.
This situation was reversed by the several new security
proofs \cite{hayashi06,hayashi07,koashi09,koashipreskill03,lo03,luo07,rennerphd,renner05,scarani08,watanabe06}, which \textbf{enabled us to study the information
reconciliation in QKD protocols without considering other
steps in QKD protocols}.

The purpose of this exposition is to introduce the problem of information
reconciliation in QKD protocols in a form accessible to coding
theorists without background in the quantum theory except footnotes and to
clarify what kind of problems arises in LDPC codes used for
information reconciliation. This exposition is organized as follows:
Section \ref{sec2} describes the problem statement and briefly reviews
the relevant research results. Section \ref{sec3} reviews the Slepian-Wolf
coding \cite[Section 15.4]{cover06} and its relation to the information
reconciliation.
Section \ref{sec4} reviews a solution by LDPC matrices
and lists the problems whose solutions are wanted
(by this author).
Section \ref{sec5} gives a conclusion.

\section{Problem statement}\label{sec2}
We assume that physical objects with two-dimensional
state spaces are transmitted in the QKD protocols.
This assumption is valid in one of several common realization of
QKD protocols. 
Another common realization of QKD protocols uses infinite-dimensional
objects \cite{grosshans03}. Information reconciliation in such a case is discussed
in \cite{bloch06,leverrier08pra,leverrier09prl,nguyen04}.
The information reconciliation in the infinite-dimensional
case seems more challenging than the two-dimensional case.

After the channel parameter estimation,
Alice has an $n$-bit binary string $X^n = (X_1$, \ldots, $X_n)$,
Bob has $Y^n = (Y_1$, \ldots, $Y_n)$,
and they know an estimate of the joint probability
distribution $P_{XY}$ assuming that $(X_i$, $Y_i)$ are
i.i.d.\  for all $i=1$, \ldots, $n$.
The goal of the information reconciliation is 
for Bob to produce a string $\hat{X}^n$ by (possibly
two-way) conversation with
Alice over the public channel.
The entire content of their conversation depends on $X^n$ and
$Y^n$, and $c(X^n, Y^n)$ denotes the entire conversation.
The desirable properties of the information reconciliation are
\begin{itemize}
\item Make $\mathrm{Pr}[X^n = \hat{X}^n]$ sufficiently close to one.
\item Make the mutual information $I(X^n ; c(X^n, Y^n))$ as small
as possible.
\end{itemize}
The reason behind the second property is that we must subtract
$I(X^n ; c(X^n, Y^n))$ bits from the length of the final secret key
\cite{hayashi07,scarani08},
because $I(X^n ; c(X^n, Y^n))$ is the amount of information
leaked to Eve during the conversation over the public channel.
Note that decreasing $I(X^n ; c(X^n, Y^n))$ is totally
different from decreasing the number of bits in the conversation
$c(X^n, Y^n)$. For example, the famous information reconciliation
protocol Cascase
\cite{brassard94,sugimoto00} exchanges many bits between
Alice and Bob, while keeping $I(X^n ; c(X^n, Y^n))$ relatively small.

We restrict ourselves to the one-way conversation, that is,
only Alice sends information to Bob and Bob sends
nothing to Alice\footnote{%
Although the Cascade \cite{brassard94,sugimoto00}
does not asymptotically yield more key,
it is also known that use of two-way conversation increases the amount
of key \cite{gottesman03,watanabe07}, which are
quantum counterparts of the two-way conversation over the public channel
proposed in
\cite{maurer93}, but we do not discuss the two-way conversation here,
because the information reconciliation with two-way conversation
seems rarely used.}.
In the one-way conversation,
$c(X^n, Y^n)$ is a function of $X^n$, denoted by $c(X^n)$.
We have $I(X^n ; c(X^n)) \leq H(c(X^n))
\leq $ the number of bits in $c(X^n)$.
We can find a good information recinciliation method by
saving the number of bits in $c(X^n)$ while
enabling Bob to decode $X^n$ from $c(X^n)$ and $Y^n$.
This is a kind of data compression problem, called
the Slepian-Wolf problem.
So we shall review it in the
next section.

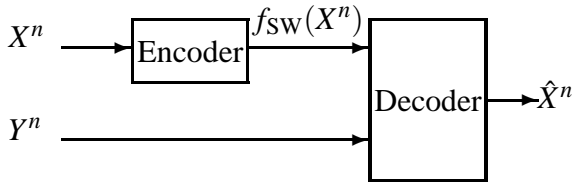
\begin{figure}[t!]
\thicklines
\begin{picture}(205,70)(65,25)
\put(60,78){$X^n$}

\put(80,78){\vector(1,0){26}}
\put(107,65){\framebox(43,23){Encoder}}
\put(152,85){$f_\mathrm{SW}(X^n)$}
\put(150,78){\vector(1,0){46}}

\put(60,43){$Y^n$}

\put(80,43){\vector(1,0){116}}

\put(197,28){\framebox(43,60){Decoder}}

\put(240,58){\vector(1,0){20}}

\put(260,55){$\hat{X}^n$}
\end{picture}
\caption{Asymmetric Slepian-Wolf coding}\label{fig1}
\end{figure}

\section{Slepian-Wolf coding}\label{sec3}
A simplified version of the general Slepian-Wolf problem \cite[Section 15.4]{cover06}
is given in Figure~\ref{fig1}.
The main information $X^n$ is statistically correlated with
the side information $Y^n$.
The encoder (data compressor) can only use $X^n$ for
generating the codeword (compressed data) $f_\mathrm{SW}(X^n)$ of
some fixed length $m$.
On the other hand, the decoder (decompresser)
can use both $f_\mathrm{SW}(X^n)$ and $Y^n$.

If $Y^n$ is unavailable by the decoder,
the compression rate $m/n$ must be $> H(X)$,
the entropy of $X^n$, in order for the decoding
error probability $\mathrm{Pr}[X \neq \hat{X}^n]$ to be
negligible.
The availability of $Y^n$ improves the optimal compression 
rate to $H(X|Y)$ from $H(X)$.
The encoder and the decoder are assumed to know (a good estimate of)
the joint probability distribution $P_{XY}$,
and they are usually optimized for a particular $P_{XY}$.
This special form of the Slepian-Wolf coding is
called \emph{asymmetric Slepian-Wolf coding} \cite{guillemot09},
because the roles of $X$ and $Y$ are asymmetric at the decoder.

We return to the information reconciliation.
Recall  that Alice has $X^n$ and Bob has $Y^n$.
If Alice sends the codeword $f_\mathrm{SW}(X^n)$,
then Bob can recover $X^n$ with high probability
by the Slepian-Wolf decoder and $Y^n$.
The amount of information leaked to Eve is estimated as
$I(X^n; f_\mathrm{SW}(X^n)) \leq H(f_\mathrm{SW}(X^n)) \leq$
the number of bits
 in $f_\mathrm{SW}(X^n)$.
Thus, if the compression rate is better\footnote{%
Strictly speaking, the use of the Slepian-Wolf coding
and the simple minimization of the number of bits in $f_\mathrm{SW}(X^n)$
neglect the optimization of the auxiliary 
random variables $U$ and $V$ in \cite{renner05}, which are
the quantum counterparts of $U$ and $Q$ in \cite[Theorem 9.2]{liang09}.},
then the upper bound on the leaked information is smaller.

\section{Use of LDPC codes and open issues}\label{sec4}
The application of LDPC codes to the Slepian-Wolf coding
with full side information can be done
as follows \cite{coleman06,liveris02}.
Let $M$ be an $m \times n$ sparse matrix, and $X^n$ be the source
information. The codeword $f_\mathrm{SW}(X^n)$ is $MX^n$.
Decoding of $X^n$ given $MX^n$ and $Y^n$ can be done by the
sum-product (belief propagation) algorithm 
over the Tanner graph of $M$.
The difference to the channel decoding by the sum-product
algorithm over the binary symmetric channels is as follows:
\begin{itemize}
\item $Y^n$ can be regarded the received word with the transmitted
word\footnote{%
It is also possible to regard that the concatenation of $X^n$ and $MX^n$ is
the transmitted word. See \cite{guillemot09} for more detail.} $X^n$ over the channel $P_{Y|X}$ with exception that
the syndrome of $X^n$ is not the zero vector but $MX^n$.
\item While the generation of messages from a check node assumes
the parity of the bits is always zero in the channel decoding,
the parity of a check node in the Slepian-Wolf decoding
is determined from $MX^n$.
\item The initial log-likelihood ratio at a variable node $X_i$
is determined from $P_{X|Y}$ and $Y_i$ in the Slepian-Wolf decoding.
\end{itemize}
Under the maximum likelihood decoding,
the sparse matrix is shown to asymptotically 
achieve the optimum compression rate \cite{muramatsu05}.
The use of sparse matrices for information reconciliation
as Slepian-Wolf encoders
seems to be first considered by Muramatsu \cite{muramatsu06}.

As a consumer of LDPC matrices for the information reconciliation,
there are at least the following problems.
\begin{enumerate}
\item\label{l1} For a given distribution $P_{XY}$, an optimized matrix $M$
is not available (on the Internet). A consumer has to find
an optimized matrix by himself using the density evolution or
its alternative.
\item\label{l2} It is convenient to have a single matrix $M$
and puncture (or shorten) $M$ for various different rates $H(X|Y)$.
\item\label{l3} For a fixed compression rate $R$, there are
infinitely many distributions $P_{XY}$ such that
$H(X|Y) = R$,
when we do not assume that $Y^n$ is the output of a binary
symmetric channel\footnote{%
Be careful that some security proofs cannot take advantage of
the nonzero difference between conditional probabilities
$P_{Y|X}(1|0)$ and $P_{Y|X}(0|1)$.
References \cite{hayashi06,hayashi07,rennerphd,renner05,scarani08}
are known to be capable of  utilizing this difference
in order to improve the compression rate in the Slepian-Wolf coding,
as pointed out in \cite[Remark 1]{watanabe08}.}
 with the input $X^n$. It is convenient to have a single $nR \times n$
matrix $M$ such that the encoder by $M$ yields small
decoding error probability with all the distributions $P_{XY}$
with $H(X|Y) \simeq R$.
\end{enumerate}

Problem \ref{l1} can be solved by a slightly modified
version of the density evolution. Under the assumption that
$Y$ is the output of a binary symmetric channel,
good sparse matrices were found by Elkouss et~al.\ 
\cite{elkouss09}. The codes in \cite{elkouss09} outperform
the Cascade \cite{brassard94,sugimoto00},
which seems the most popular method for the
information reconciliation in QKD protocols
when this exposition is written.
Thus, the use of LDPC matrices looks promising
for QKD protocols.

Problems \ref{l2} and \ref{l3} are large disadvantages
compared to the Cascade \cite{brassard94,sugimoto00},
because the Cascade is in a sense universal and
we do not have to adjust it to different $P_{XY}$.
In order for the LDPC method to become more favorable as an alternative
to the Cascade in the QKD application,
these problems may have to be solved.

Problem \ref{l2} was considered by Varodayan et~al.\ 
\cite{varodayan06}, in which an accumulator is serially
connected to an LDPC encoder. However, the performance
is still a bit distant from the theoretical optimum, and
there seems to be a room for improvement.
Several other solutions have been proposed and can be found in
\cite{guillemot09}.

Although Coleman \cite{colemanphd} provided a Shannon theoretic solution
to Problem \ref{l3} with the expander code and the
minimum entropy decoder by the linear programing,
an efficient solution has not been provided as far as the
author knows.

\section{Conclusion}\label{sec5}
The standard error-correction scheme, such as LDPC codes
and turbo codes, seems less popular than the Cascade
protocol \cite{brassard94,sugimoto00} for the information reconciliation in quantum
key distribution protocols. The author guessed the difficulty
in selecting optimized codes as the reason
for its unpopularity, and gave three specific difficulties.

\section*{Acknowledgment}
The author would like to thank Dr.\ Manabu Hagiwara
for helpful comments on an earlier manuscript, Mr.\ Anthony Leverrier
for drawing his attention to  \cite{leverrier08pra,leverrier09prl}, and
Mr.\ Tetsunao Matsuta for
providing Fig.\ \ref{fig1}.


\urlstyle{rm}

\end{document}